# Advanced bridge instrument for the measurement of the phase noise and of the short-term frequency stability of ultra-stable quartz resonators

F. Sthal, X. Vacheret, S. Galliou
P. Salzenstein, E. Rubiola
FEMTO-ST Institute, UMR CNRS 6174
Besançon, France
fsthal@ens2m.fr

G. Cibiel
Microwave and Time-Frequency Department
CNES
Toulouse, France
gilles.cibiel@cnes.fr

*Abstract* — **This article reports on the advances in the characterization of an instrument intended to test 5 MHz and 10 MHz crystal resonators packaged in HC40 enclosure, that used for the high-stability space resonators. Improvements concern the design of new double ovens, the reduction of the residual carrier, a lower drift, and a modified calibration process. The instrument sensitivity, that is, the background phase noise converted into Allan deviation, is of $10^{-14}$.**

I. INTRODUCTION

High-stability quartz oscillators are needed in a number of space applications. A short-term stability of parts in $10^{-14}$ [Allan deviation $\sigma_y(\tau)$] is sometimes required, for integration time $\tau$ of approximately 1–10 s. The Centre National d'Etudes Spatiales (CNES) and the FEMTO-ST Institute have been collaborating for many years in this domain, aiming at measuring and at understanding the oscillator noise [1]. The highest stability has been observed on 5 MHz and 10 MHz bulk acoustic-wave resonators. Yet this stability is still not sufficient, or the manufacturing method is not reproducible. Recently, the analysis of a few premium-stability oscillators has demonstrated that the oscillator frequency instability is due to the fluctuation of the resonator natural frequency, rather than to the noise of the sustaining amplifier via the Leeson effect [2]. It is therefore natural to give attention to the measurement of the resonator fluctuations.

This paper presents improvements of an instrument intended to test 5 MHz and 10 MHz high-stability resonators packaged in HC40 enclosure that used in ultra stable oscillators. The instrument is based on some known ideas, namely: (1) the frequency stability is measured through the phase noise spectral density $S_\varphi(f)$, converted into $S_y(f)$, and then into $\sigma_y(\tau)$; and (2) $S_\varphi(f)$ is measured with the bridge (interferometric) scheme. This scheme provides the lowest background noise, and enables the measurement at low drive level. Improvements concern the design of new double ovens, the reduction of the residual carrier, a lower drift, and a modified calibration process.

II. BASIC PRINCIPLE ON MEASUREMENT BENCH

Fig. 1 shows the principle of resonator phase noise measurement bench using carrier suppression technique. The carrier signal of the driving source is split into two equal parts to drive both devices under test (DUT). The DUTs can be resistors (to measure the noise floor of the system) or crystal resonator pairs.

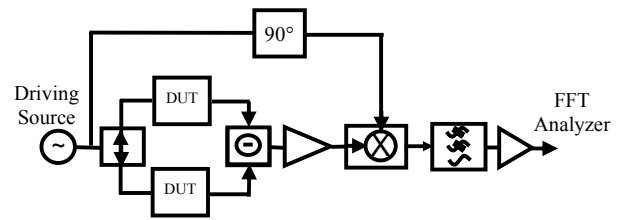

Figure 1. Principle of the measurement bench.

The resonant frequency of each arm of the bridge is tuned to the driving source frequency with a series tuning capacitor. The carrier signal is canceled when both signals are combined 180° out of phase. Since phase noise is defined relative to the carrier power, reducing the carrier has the effect of amplifying the phase noise of the DUT. The output signal is amplified and then detected by the phase noise detector.

Calibration of the measurement system is obtained by injecting a known side band on one of the arms of the bridge. The noise of the DUT, as seen on the fast Fourier transform analyzer, is corrected using the calibration factor determined with the side band.

Fig.2 shows the measurement system. The new bench developed here has been obtained from two previous systems used in our laboratory experiences [3-4].

Boxes concept gives a lot of facilities to build the system.

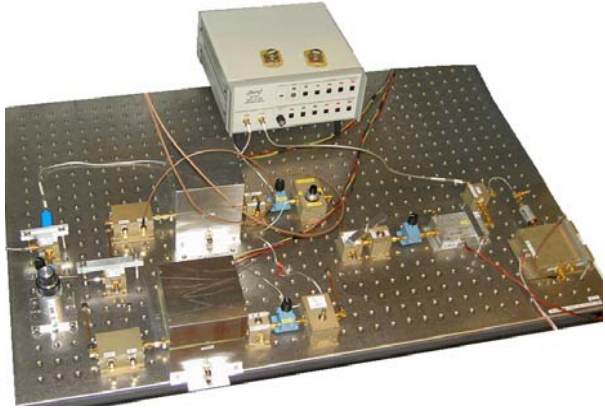

Figure 2.  Resonator phase noise measurement bench.

Any measurement requires a data acquisition.. A specific acquisition interface has been developed. It is based on HP3561A FFT analyzer. The main interface window is presented in Fig. 3. Frequency range and numbers of average can be well adapted according to the measurement time.

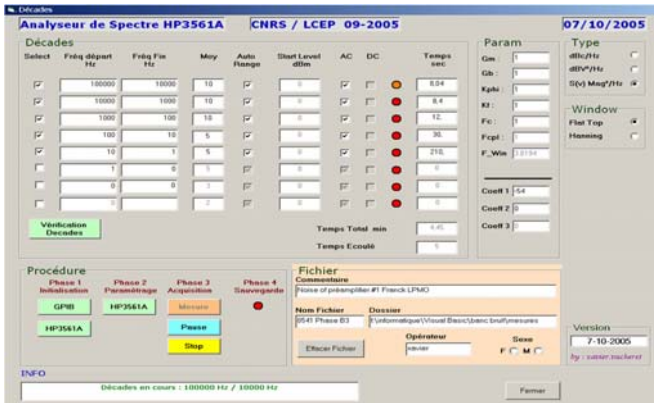

Figure 3.  Data acquisition interface.

III. IMPROVEMENTS

*A. Resonator ovens*

The main improvement of ovens is shown in Fig. 4. Fig. 4a shows the crystal box of the new ovens. It consists of a double enclosure and two thermally controlled ovens, used in order to control the quartz crystal temperature. The temperature of the crystal oven is controlled by means of a tuning resistor chip. A resistor step of about 30 Ω allows a temperature adjustment of 0.05 °C in the 60 to 90 °C temperature range. In our new crystal ovens, the tuning capacitors of the quartz crystals are inserted in the crystal oven. Fig. 4b shows the crystal oven and the inserted location of the tuning capacitor. Fig. 4c shows the cable ovens.

By this way, a remote temperature control can be easily achieved without opening the overall device. The cable oven temperature is set at a fixed value between the room temperature and the crystal temperature.

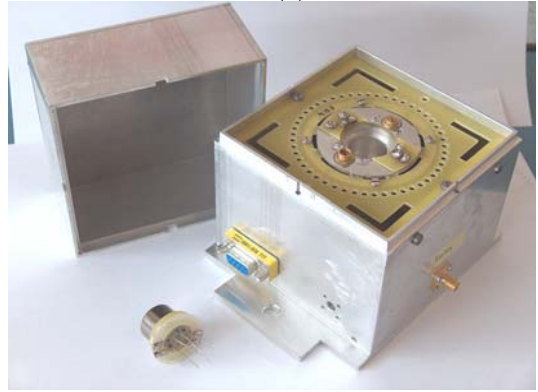
(a)

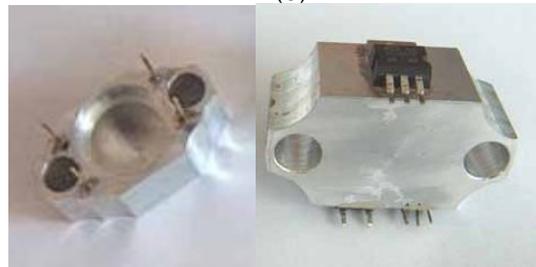
(b)

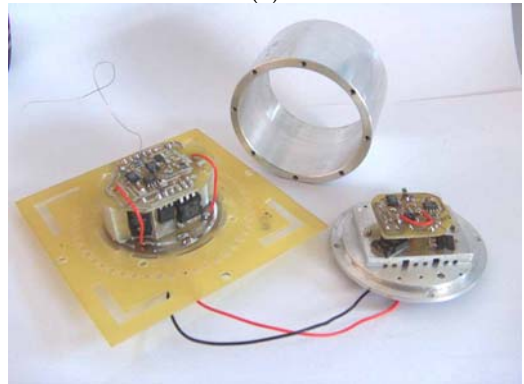
(c)

Figure 4.  Double quartz crystal ovens.
(a) oven top view, (b) crystal oven, (c) cables oven.

Performances of these crystal ovens have been measured according to the procedure describe in a previous paper [5]. The thermal stability is about 2 µ°C at 1 s and about 20 µ°C at 10 s. The relative frequency fluctuation of $\Delta f/f$ of the bench due to the thermal effects on the resonator is about $2 \cdot 10^{-15}$ at 1 s.

*B. Temperature effect on tuning capacitors*

The crystal tuning capacitor, located in the enclosure of quartz, is subjected to the fluctuations of temperature brought by the thermal regulation of the thermostat. The capacitor varies according to a thermal coefficient. In our case, standard variable capacities 5600 Johanson having a temperature coefficient of ±30 ppm/°C are used. Quartz crystal and its tuning capacitor are represented in Fig. 5.

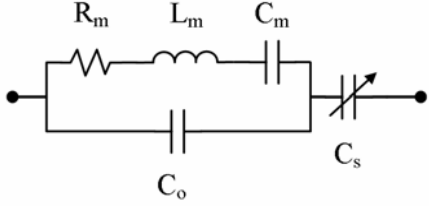

Figure 5. Quartz crystal model and its serial tuning capacitor.

$R_m$, $L_m$ and $C_m$ are the motional parameters of the quartz crystal and $C_0$ is the static capacitor. Typical equivalent parameters of 5 MHz quartz crystal resonator are:
Quartz crystal static capacitor: $C_0 \approx 3$ pF
Quartz crystal motional capacitor: $C_m \approx 0.2$ fF

The relative variation of the frequency is classically given by the following equation:

$$\frac{\Delta f}{f_0} \approx \frac{C_m}{2(C_0 + C_s)} \quad (1)$$

With $C_s$: Serial tuning capacitor associated to the quartz crystal and $f_0$ the resonant frequency of the resonator.

Thus, the resonant frequency of the resonator and its tuning capacitor is given by:

$$f = f_0 + \Delta f = f_0\left(1 + \frac{\Delta f}{f_0}\right) = f_0 \times \left(1 + \frac{C_m}{2(C_0 + C_s)}\right) \quad (2)$$

The fluctuations $\Delta C_s$ of $C_s$ due to the thermal fluctuations give a relative frequency variation expressed by:

$$\frac{\Delta f}{f_0} = -\left(\frac{C_m}{2(C_0 + C_s)^2}\right) \times C_s \times \left(\frac{\Delta C_s}{C_s}\right) \quad (3)$$

Fig. 6 shows the relative frequency fluctuations according to the tuning capacitor $C_s$ for a temperature variation of 1 °C.

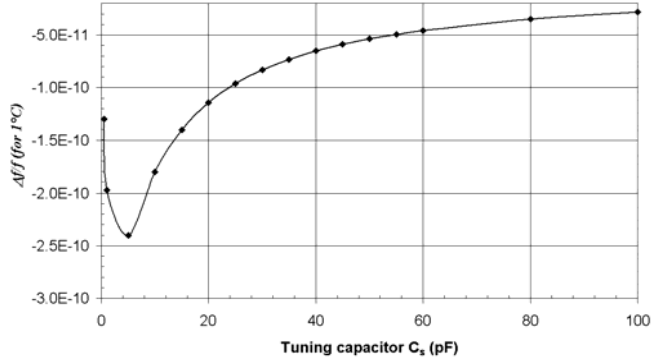

Figure 6. $\Delta f/f$ due to the tuning capacitor $C_s$ for 1°C.

In the worst case, for $C_s = 2.9$ pF, the relative frequency fluctuations are about $2.4 \cdot 10^{-10}$ for a temperature variation of 1 °C. According to the thermal stability of the ovens at 10 s, the relative frequency fluctuation of $\Delta f/f$ of the bench due to the thermal effects on the tuning capacitor will not be above $5 \cdot 10^{-15}$. This oven configuration is well adapted to measure resonators with an Allan standard deviation stability around $1 \cdot 10^{-14}$.

## C. Calibration

Calibration of the measurement system is obtained by injecting a known side band on one arm of the bench. This sideband simulates an equivalent phase noise. When quartz crystal resonators are measured, the calibration sideband must be injected inside the bandwidth of the resonator. Fig. 7 shows the FFT level obtained from the known sideband versus the difference between the source frequency and the sideband frequency. Values are given according to the powers dissipated by the quartz crystal.

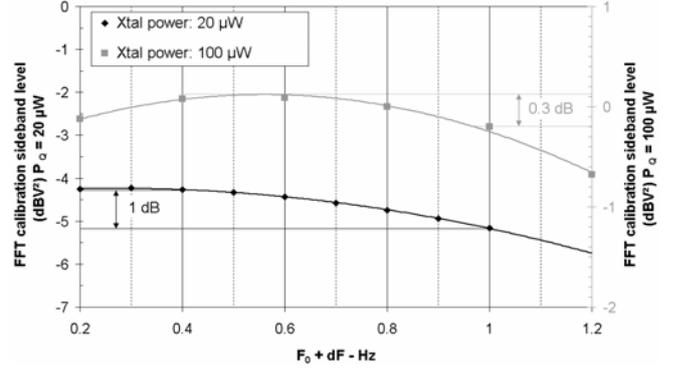

Figure 7. Fluctuation of the calibration sideband level obtnained at the FFT analyzer output in dBV² according to the frequency offset between the carrier frequency and the sideband frequency.

The frequency of the calibration sideband is moved from 0.2 Hz to 1.2 Hz from the carrier frequency. The maximum of the curves is located towards a frequency offset of 0.4 Hz. The level difference between 0.4 Hz and 1 Hz can vary between 0.3 dB and 1 dB. The calibration side band is very important. Indeed, the measured noise is directly modified by the correction factor as shown in Fig. 8.

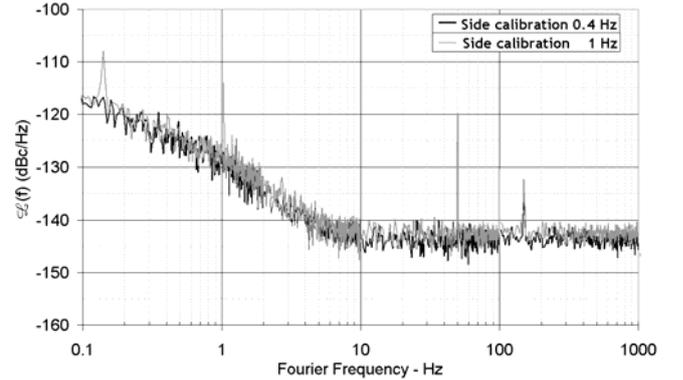

Figure 8. Calibration side band effect on phase noise measurment of 5 MHz quartz crystal resoantor (crystal dissipated power: 20 µW).

Table 1 gives the corresponding fluctuations in term of Allan standard deviation $\sigma_y$.

TABLE I. $\sigma_Y$ VARIATION ACCORDING TO SIDEBAND CALIBRATION ERROR (QUARTZ CRSYTAL POWER: 20 µW).

| | |
|---|---|
| $\sigma_y(\tau)$ – calibration side band 0.4 Hz | $1.3 \cdot 10^{-13}$ |
| $\sigma_y(\tau)$ – calibration side band 1 Hz | $1.8 \cdot 10^{-13}$ |

The calibration sideband must be to the maximum amplitude of quartz crystal transfer function to not generate an error of measurement and increase the uncertainties of measurements. The value of the calibration sideband was thus fixed at 0.4 Hz from the source frequency instead of below 2 Hz.

### D. Output offset

The FFT analyzer input is sensitive to the DC voltage. The calibration range of the analyzer can become inappropriate if this voltage fluctuate during the measures. Before the measurement, the delay line of the measurement bench is tuned to give a signal at the phase detector output with no DC voltage. Unfortunately, the experiment duration to get low frequency (about 1 hour for 0.01Hz) shows us that the output offset can drift.

Better understanding of this parameter has been obtained by measuring its variations. For that, we used the FFT HP3621A analyzer which is able to measure the DC residual voltage. An interface allowing a continuous acquisition of the offset was developed (Fig. 9). The output offset is measured during an integration time of 2 seconds. Its evolution is observed with a space time 5 seconds interval between each measurement. The observation time can be higher than 7 days.

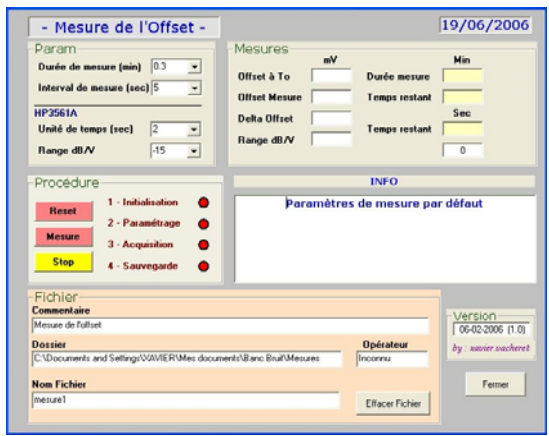

Figure 9. Offset HP3561A acquisition window.

Initially, we examined the offset stability of the DC-100 kHz amplifier used after the phase detector. Its offset variations are below 2 mV which is remarkably stable.

The second analysis consists in measuring the offset during the phase of adjustment in a simple arm of the bench. Resistor is used to measure the noise floor of the system. The delay line is tuned to have the simple arm and the pump signal 90° out of phase. Fig. 10 shows the offset measurements with a dissipated power in resistor equal to 75 µW. Curves are given for a time up to 2 hours which is above classically phase noise measurement.

A first problem is revealed by these measurements. The nature of the connection between the element to be measured and its support has been examined. It is easier to use a mechanical contact to facilitate the assembly and disassembly of the DUT that must be measured. Commercial benches often use this kind of support. Unfortunately, saved time is with the detriment of the stability of the system and the quality of measurement carried out. It is preferable in this case to weld the connections between the DUT (resonator) and the support.

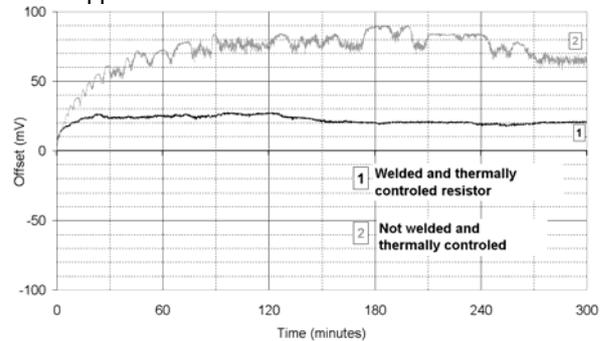

Figure 10. Offset stability according to the connection nature between DUTs and support.

Another series of test have been achieved in order to limit the offset drift during the measurement time. An answer is come from the attenuator-phase converter units of each arm. These units are used to adjust the carrier suppression. Fig. 11 shows the offset drift for two kind of attenuator. The 0-20 dB attenuators were used in our previous version of our bench. Usual adjustment in term of amplitude level is always around 3 dB. We thus replaced the 0-20 dB attenuator by a 0-8 dB attenuator. This enables us to still have a comfortable margin of adjustment and to decrease the effect of the thermal relaxation by 60% what is a considerable profit is appreciable taking into consideration expected performance.

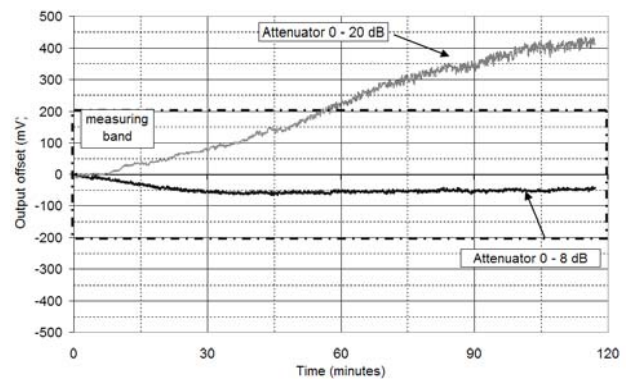

Figure 11. Offset drift according different attenuators.

Offset variations are a good indicator of reliability of measurements. A null offset indicates a state of quasi-perfect balance between both arms of the bench. It is necessary to seek a limit point to consider that the measurement is not affected by the offset. Fig. 12 shows 5 MHz quartz crystal resonator phase noise measurements with a voluntary "degraded" value of offset (about 200 mV) compared with a measurement having an offset adjustment. Insignificant differences are observed between both curves. Thus, an offset below 200 mV warrants a good measurement.

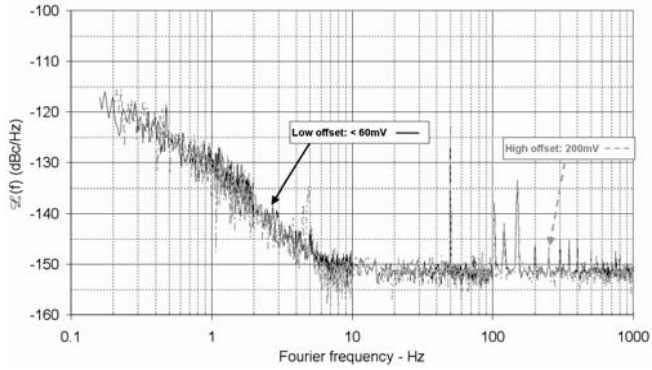

Figure 12. Quartz crystal phase noise measurement according to ouptut offset Quartz crystal dissipated power (100 μW).

*E. Carrier suppression*

The carrier suppression is a critical phase in the bench adjustment. It depends on many factors (dissipated power, phase tuning, stabilization time, measurement environment) that can interact on stability, adjustment facilities or attenuation level. If the ideal adjustment exists in theory, in practice, the total carrier suppression without fluctuations during a long time is really difficult to obtain, even impossible to realize. We carried out comparative measurements on power levels and carrier suppression. We have sought a range of adjustment in which a reliable and reproducible measurement is guarantee. A limit was established concerning the minimal value of the carrier suppression level. Table II gives carrier suppression level that can be acceptable to get a good phase noise measurement.

TABLE II. ACCEPTABLE CARRIER SUPPRESSION LEVEL.

| Minimum value | –65 dB |
|---|---|
| **Accepted Minimum value** | **–75 dB** |
| Mean value | –82 dB |
| Maximum value | –98 dB |

Fig. 13 shows the measurement of a quartz crystal pair with both different carrier suppression levels. A same behavior is observed according to the power applied on the resonator.

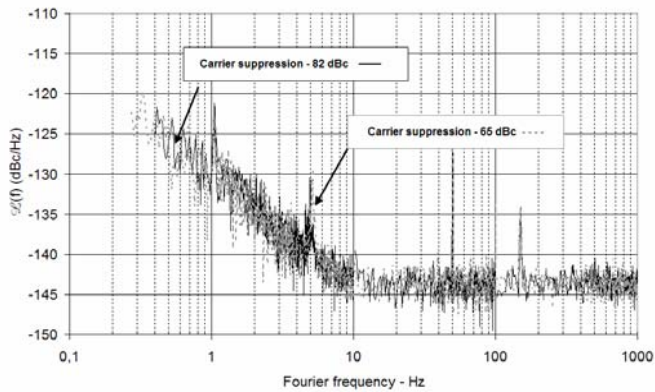

Figure 13. Effect of the carrier supression level on phase noise measurement of a quartz crystal pair ($P_Q$ = 20 μW).

It should be noted that in practice the average carrier suppression level is in the neighbourhoods of - 82 dB. According to many measurements that we have done, a safety margin of about - 10 dB can be considered.

*F. Measurement environment*

Influence of the environment to the phase noise measurement is illustrated in Fig. 14. Undesirable peak appeared around 7 Hz. This sporadic and random phenomenon has been observed several times. It is a student festival of music which revealed us the sensitivity of our device to the low-frequency vibrations. The loudness and the test periods of the loudspeakers allow us to establish the link between the effect and the cause. The measurement frequencies to get the phase noise of the resonator are classically in the bandwidth 0.01 Hz to 10 Hz. This frequency range is particularly sensitive to mechanical vibrations. Vibration isolation tables must be chosen very carefully regarding their cut-off frequencies which are in this frequency band. With active control tables, it is also possible to amplify parasitic signals inside the bandwidth. This problem must be taking into account very carefully. After several tests, localization of the measurement bench was chosen according to this phenomenon.

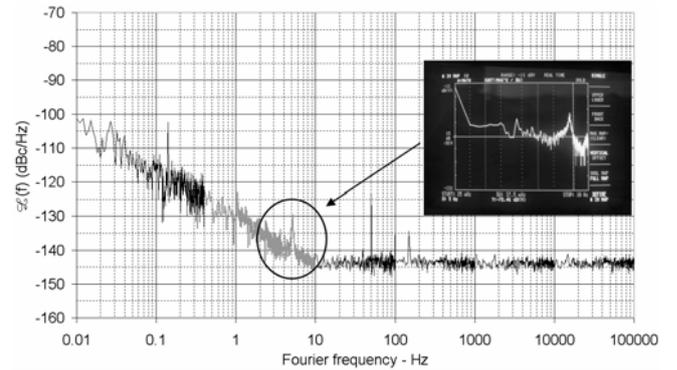

Figure 14. Effect of the measurement environment.

*G. Trend curves*

Resonator noise is measured through the phase noise spectral density and then converted into Allan standard deviation. The Allan variance is computed from the Fourier frequency and from the power spectral density of the frequency fluctuations [6]:

$$\sigma_y^2(\tau) = \int_0^\infty S_y(f) \frac{2\sin^4 \pi f \tau}{(\pi f \tau)^2} df \quad (4)$$

with $f$ = Fourier frequency. The standard Allan variance can not be calculated from the phase noise of the resonator unless one knows the loaded $Q_L$ of the resonator. The equations below show how $\sigma_y(\tau)$ is calculated for the flicker phase noise in a resonator. The resonator is considered as a low pass filter. Thus, we have:

$$S_y(f) = \left(\frac{1}{2Q_L}\right)^2 S_\phi(f) \quad (5)$$

Here $S_\phi(f)$ is the measured phase noise of the resonator, $Q_L$ is the loaded quality factor of the resonator and $S_y(f)$ is the power spectral density of the frequency fluctuations of the resonator. The flicker phase in $S_\phi(f)$ gives flicker frequency in $S_y(f)$.

As expected $\sigma_y(\tau)$ is independent of $\tau$ for flicker frequency noise, $\sigma_y(\tau)$ is:

$$\sigma_y^2(\tau) = 2\ln(2) S_y(f=1Hz) \quad (6)$$

Thus the Allan deviation is:

$$\sigma_y^2(\tau) = 2\ln(2)\left(\frac{1}{2Q_L}\right)^2 S_\phi(f=1Hz) \quad (7)$$

The Allan deviation calculated this way represents the Allan deviation of an oscillator containing the test resonator in which the only source of flicker frequency noise is the test resonator.

The loaded quality factor $Q_L$ is given by means of the Lesson cutoff frequency which is equal to:

$$f_L = \frac{v_0}{2Q_L} \quad (8)$$

The cut-off frequency is traditionally obtained graphically. This treatment is often subject to interpretation which can vary from one operator to another. To avoid dispersion, we propose a calculation approach by trend curves with $f^{-1}$ and $f^{-3}$ slopes (Fig.15). These slopes correspond to the low-pass filter effect of the frequency flicker resonator noise.

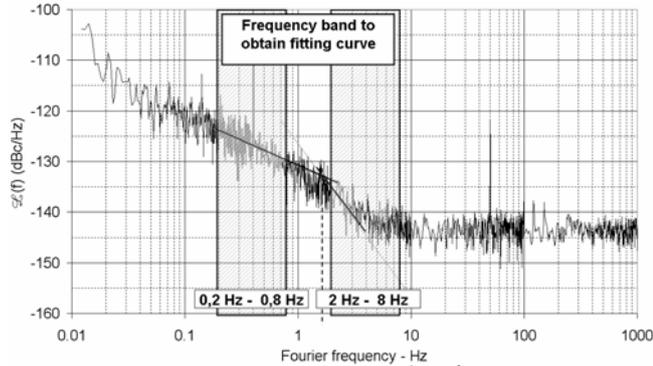
Figure 15. Frequency bandwidth for $f^{-1}$ et $f^{-3}$ interpolation.

The ranges of the frequencies are chosen according to the slope variations. This method can be applied directly in the phase noise measurements (Fig. 16). A better choice can be obtained by injecting a known white noise at resonator input and observe its low-pass filter effect (Fig. 17). The curve corresponds to the resonator transfer function. In this last case, due to the level of the known noise source, the frequency ranges to get the trend curves are more important and give more precision in cutoff frequency determination.

Classically, for 5 MHz quartz crystal resonator, the frequency range of the $f^0$ slope is 0.2 Hz-0.8 Hz and the frequency range of the $f^{-2}$ slope is 2 Hz-6 Hz.

Values of $\sigma_y$ are compared in Table 3 and 4. Both measured resonators are considered identical that means $S_\phi(f)$ is equal to the single sideband power spectral density of phase fluctuations $\mathscr{L}(f)$. A significant difference can be observed in term of cut-off frequency. A precision of about 0.1 Hz is obtained on the Leeson frequency. According to the analyzer, $\mathscr{L}(f)$ can be measured with ± 2 dB of precision.

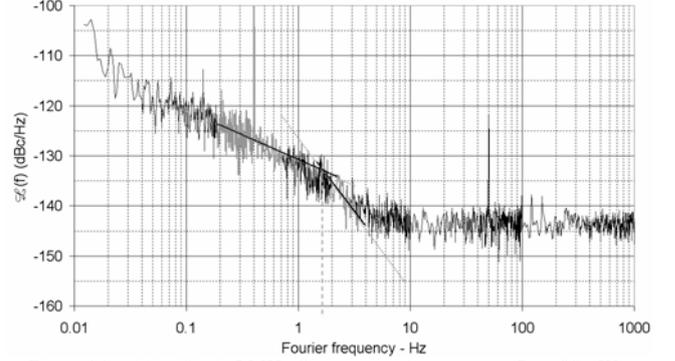
Figure 16. $\mathscr{L}(f)$ of both 5 MHz Quartz crystal resonators ($P_Q = 20\ \mu W$).

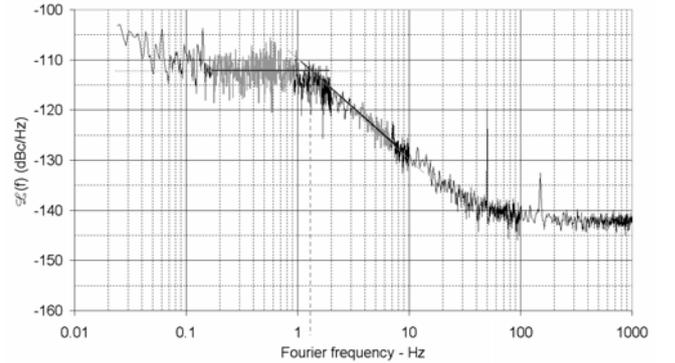
Figure 17. Measurement of the resonator transfer function.

TABLE III. ALLAN STANDARD DEVIATION OBTAINED WITH $f_L$ FROM FIG. 16.

|  | $\sigma_y$ | $\mathscr{L}$(1 Hz) | | |
|---|---|---|---|---|
|  |  | -130 | -132 | -134 |
|  | 1,6 | 1,19E-13 | 9,46E-14 | 7,52E-14 |
| $f_L$ | 1,7 | 1,27E-13 | **1,01E-13** | 7,99E-14 |
|  | 1,8 | 1,34E-13 | 1,06E-13 | 8,46E-14 |

TABLE IV. ALLAN STANDARD DEVIATION OBTAINED WITH $f_L$ FROM FIG. 17.

|  | $\sigma_y$ | $\mathscr{L}$(1 Hz) | | |
|---|---|---|---|---|
|  |  | -130 | -132 | -134 |
|  | 1,3 | 9,68E-14 | 7,69E-14 | 6,11E-14 |
| $f_L$ | 1,4 | 1,04E-13 | **8,28E-14** | 6,58E-14 |
|  | 1,5 | 1,12E-13 | 8,87E-14 | 7,05E-14 |

Finally, $\sigma_y$ is given by two extreme values. This method is reproducible and allows the same treatment in a resonator batch. Comparison between resonators is really simplified.

## IV. CONCLUSION

New double ovens have been designed to improve the ultimate noise floor of our carrier suppression bench. These new crystal ovens present an Allan standard deviation of about $2 \cdot 10^{-15}$ at 1 s in terms of relative frequency fluctuations.

Several improvements have been presented in order to increase the reliability of the phase noise measurement bench of 5 MHz and 10 MHz crystal resonators. The developed bench is now suitable to measure manufacturing production of resonators using this passive method.

## ACKNOWLEDGMENT

Authors thank CNES for the funding support.